\documentstyle[pra,aps,multicol]{revtex}
\renewcommand{\narrowtext}{\begin{multicols}{2} \global\columnwidth20.5pc}
\renewcommand{\widetext}{\end{multicols} \global\columnwidth42.5pc}
\begin{document}
\draft
\title{{\bf Calculating the relative entropy of entanglement}}
\author{Shengjun Wu and Yongde Zhang}
\address{{\it Department of Modern Physics,}\\
University of Science and Technology of China, Hefei, Anhui 230027,\\
P.R.China}
\date{\today}
\maketitle

\begin{abstract}
We extend Vedral and Plenio's theorem (theorem 3 in Phys. Rev. A 57, 1619)
to a more general case, and obtain the relative entropy of entanglement for
a class of mixed states, this result can also follow from Rains' theorem
(theorem 9 in Phys. Rev. A 60, 179).
\end{abstract}


\vskip 1.0cm

\narrowtext

The relative entropy of entanglement plays an important role in
understanding quantum entanglement \cite{vprk}, it reduces to the von
Neumann reduced entropy of either side for the bipartite pure states \cite
{vp}; for general mixed states, the relative entropy of entanglement is hard
to calculate. In this note, we extend Vedral and Plenio's result for pure
states to a more general case, and calculate the relative entropy of
entanglement for a class of mixed states.

The relative entropy of entanglement for 2-party (say, A and B) quantum
state $\sigma $ is defined by \cite{vprk}:

\begin{equation}
Er(\sigma )\equiv {\min }_{\rho \in D}S(\sigma ||\rho )  \label{a0}
\end{equation}
where $D$ is the set of all disentangled states, and $S(\sigma ||\rho
)\equiv tr\{\sigma (\ln \sigma -\ln \rho )\}$ is the relative entropy of $%
\sigma $ with respect to $\rho $. Let $\rho ^{*}$ denotes the disentangled
state that minimizes the relative entropy $S(\sigma ||\rho )$. In fact, to
calculate $Er(\sigma )$ is to find the state $\rho ^{*}$.

{\sl Main result:} For bipartite quantum state 
\begin{eqnarray}
\sigma  &=&{\sum }_{n_1n_2}a_{n_1n_2}\left| \phi _{n_1}\psi
_{n_1}\right\rangle \left\langle \phi _{n_2}\psi _{n_2}\right|   \nonumber \\
&=&{\sum }_{n_1n_2}a_{n_1n_2}\left| \phi _{n_1}\right\rangle \left\langle
\phi _{n_2}\right| \otimes \left| \psi _{n_1}\right\rangle \left\langle \psi
_{n_2}\right|   \label{a}
\end{eqnarray}
the relative entropy of entanglement is given by 
\begin{equation}
Er(\sigma )=-{\sum }_na_{nn}\ln a_{nn}-S(\sigma )  \label{b}
\end{equation}
and the disentangled state that minimizes the relative entropy is 
\begin{equation}
\rho ^{*}={\sum }_na_{nn}\left| \phi _n\psi _n\right\rangle \left\langle
\phi _n\psi _n\right|   \label{c}
\end{equation}
Here, $\left| \phi _n\right\rangle $ ($\left| \psi _n\right\rangle $) is a
set of orthogonal normalized states of system A (B); $S(\sigma )$ is the von
Neumann entropy defined by 
\begin{equation}
S(\sigma )\equiv tr\{-\sigma \ln \sigma \}  \label{c1}
\end{equation}

If $\sigma $ is a pure state, there is $a_{n_1n_2}=\sqrt{p_{n_1}p_{n_2}}$,
our result reduces to Vedral and Plenio's result.

{\sl Proof.} The proof is similar to the proof of theorem 3 in \cite{vp}.
Since we already have a guess for $\rho ^{*}$, we show that the gradient $%
\frac d{dx}S(\sigma ||(1-x)\rho ^{*}+x\rho )|_{x=0}$ for any $\rho \in D$ is
non-negative. On the other hand, if $\rho ^{*}$ was not a minimum, the above
gradient would be strictly negative, which is a contradiction. Thus the
result follows.

Let $f(x,\rho )\equiv S(\sigma ||(1-x)\rho ^{*}+x\rho )$, using the identity 
$\ln A=\int_0^\infty [(At-1)/(A+t)]dt/(1+t^2)$, we can get 
\begin{equation}
\frac{\partial f}{\partial x}(0,\rho )=1-\int_0^\infty tr[(\rho
^{*}+t)^{-1}\sigma (\rho ^{*}+t)^{-1}\rho ]dt  \label{d}
\end{equation}
Since $\rho ^{*}={\sum }_{n}a_{nn}\left| \phi _n\psi _n\right\rangle
\left\langle \phi _n\psi _n\right| $, it is not difficult to get 
\begin{eqnarray}
&&(\rho ^{*}+t)^{-1}\sigma (\rho ^{*}+t)^{-1}  \nonumber \\
&&={\sum }_{nn^{^{\prime }}}(a_{nn}+t)^{-1}\cdot a_{nn^{^{\prime }}}\cdot
(a_{n^{^{\prime }}n^{^{\prime }}}+t)^{-1}\left| \phi _n\psi _n\right\rangle
\left\langle \phi _{n^{^{\prime }}}\psi _{n^{^{\prime }}}\right|  \label{e}
\end{eqnarray}
Set $g(n,n^{^{\prime }})\equiv a_{nn^{^{\prime }}}\cdot \int_0^\infty
(a_{nn}+t)^{-1}\cdot (a_{n^{^{\prime }}n^{^{\prime }}}+t)^{-1}dt$, obviously
we have that $g(n,n)=1$, and for $n\neq n^{^{\prime }}$%
\begin{equation}
g(n,n^{^{\prime }})=a_{nn^{^{\prime }}}\cdot \frac{\ln a_{nn}-\ln
a_{n^{^{\prime }}n^{^{\prime }}}}{a_{nn}-a_{n^{^{\prime }}n^{^{\prime }}}}
\label{f}
\end{equation}

We now show that $|g(n,n^{^{\prime }})|\leq 1$. Ref. \cite{vp} has proved
that 
\begin{equation}
0\leq \sqrt{a_{nn}a_{n^{^{\prime }}n^{^{\prime }}}}\cdot \frac{\ln
a_{nn}-\ln a_{n^{^{\prime }}n^{^{\prime }}}}{a_{nn}-a_{n^{^{\prime
}}n^{^{\prime }}}}\leq 1  \label{g}
\end{equation}
so we only need to show that $|a_{nn^{^{\prime }}}|\leq \sqrt{%
a_{nn}a_{n^{^{\prime }}n^{^{\prime }}}}$. Let $\left| \Psi \right\rangle
=a\left| \phi _n\psi _n\right\rangle +b\left| \phi _{n^{^{\prime }}}\psi
_{n^{^{\prime }}}\right\rangle $, $a$ and $b$ are arbitrary complex numbers.
Since $\sigma $ is a quantum state, it follows that 
\begin{equation}
\left\langle \Psi \right| \sigma \left| \Psi \right\rangle \geq 0  \label{g1}
\end{equation}
for arbitrary pair of $a$ and $b$, this requires 
\begin{equation}
a_{nn}a_{n^{^{\prime }}n^{^{\prime }}}-a_{nn^{^{\prime }}}a_{n^{^{\prime
}}n}=a_{nn}a_{n^{^{\prime }}n^{^{\prime }}}-|a_{nn^{^{\prime }}}|^2\geq 0
\label{i}
\end{equation}
Therefore 
\begin{equation}
|g(n,n^{^{\prime }})|\leq 1  \label{j}
\end{equation}

The following steps are just the same as those in ref. \cite{vp}, which are
written here for the completeness of this proof.

Let $\rho \equiv \left| \alpha \right\rangle \left\langle \alpha \right|
\otimes \left| \beta \right\rangle \left\langle \beta \right| $ where $%
\left| \alpha \right\rangle =_{n}{\sum }a_n\left| \phi _n\right\rangle $ and 
$\left| \beta \right\rangle =_{n}{\sum }b_n\left| \psi _n\right\rangle $ are
normalized vectors, it is not difficult to show that 
\begin{equation}
\frac{\partial f}{\partial x}(0,\rho )-1=-{\sum }_{n_1n_2}g(n_1,n_2)\cdot
a_{n_2}b_{n_2}a_{n_1}^{*}b_{n_1}^{*}  \label{k}
\end{equation}
therefore 
\begin{eqnarray}
\left| \frac{\partial f}{\partial x}(0,\rho )-1\right| & &  \nonumber \\
\leq& &{\sum }_{n_1n_2}|g(n_1,n_2)|\cdot |a_{n_2}|\cdot |b_{n_2}|\cdot
|a_{n_1}^{*}|\cdot |b_{n_1}^{*}|  \nonumber \\
\leq& &{\sum }_{n_1n_2}|a_{n_2}|\cdot |b_{n_2}|\cdot |a_{n_1}^{*}|\cdot
|b_{n_1}^{*}|  \nonumber \\
=&&({\sum }_{n}|a_n|\cdot |b_n|)^2  \nonumber \\
\leq &&{\sum }_{n}|a_n|^2\cdot {\sum }_{n}|b_n|^2  \nonumber \\
=&&1  \label{l}
\end{eqnarray}
Then it follows that $\frac{\partial f}{\partial x}(0,\left| \alpha \beta
\right\rangle \left\langle \alpha \beta \right| )\geq 0$. Since any
disentangled state $\rho \in D$ can be written in the form $\rho ={\sum }%
_{i}r_i\left| \alpha ^i\beta ^i\right\rangle \left\langle \alpha ^i\beta
^i\right| $, we have that 
\begin{equation}
\frac{\partial f}{\partial x}(0,\rho )={\sum }_{i}r_i\cdot \frac{\partial f}{%
\partial x}(0,\left| \alpha ^i\beta ^i\right\rangle \left\langle \alpha
^i\beta ^i\right| )\geq 0  \label{m}
\end{equation}

Now we show that $S(\sigma ||\rho )\geq S(\sigma ||\rho ^{*})$ for all $\rho
\in D$. Suppose that $S(\sigma ||\rho )<S(\sigma ||\rho ^{*})$ for some $%
\rho \in D$, then, for $0<x\leq 1$, 
\begin{eqnarray}
f(x,\rho ) &=&S(\sigma ||(1-x)\rho ^{*}+x\rho )  \nonumber \\
&\leq& (1-x)S(\sigma ||\rho ^{*})+xS(\sigma ||\rho )  \label{n} \\
&=&(1-x)f(0,\rho )+xf(1,\rho )  \nonumber
\end{eqnarray}
which implies 
\begin{equation}
\frac{f(x,\rho )-f(0,\rho )}x\leq f(1,\rho )-f(0,\rho )<0  \label{o}
\end{equation}
This contradicts the fact that $\frac{\partial f}{\partial x}(0,\rho )\geq 0$
in the limit of small $x$. So, for any $\rho \in D$, this is $S(\sigma
||\rho )\geq S(\sigma ||\rho ^{*})$, i.e., the state $\rho ^{*}= {\sum }%
_{n}a_{nn}\left| \phi _n\psi _n\right\rangle \left\langle \phi _n\psi
_n\right| $ minimizes the relative entropy $S(\sigma ||\rho )$ over $\rho
\in D$.

Then it follows that 
\begin{equation}
Er(\sigma )=tr\{\sigma (\ln \sigma -\ln \rho ^{*})\}=-{\sum }_{n}a_{nn}\ln
a_{nn}-S(\sigma )  \label{p}
\end{equation}
This completes the proof.

It should be pointed out that the above result can also follow directly from
Rains' theorem 9 in ref. \cite{rains}.

The Vedral and Plenio's theorem has been extended to a more general case,
the result is, nevertheless, very useful for calculating the relative
entropy of entanglement for the mixed states $\sigma ={\sum }%
_{n_1n_2}a_{n_1n_2}\left| \phi _{n_1}\psi _{n_1}\right\rangle \left\langle
\phi _{n_2}\psi _{n_2}\right| $. For the simplest case of two qubits, the
relative entropy of entanglement $Er$ of the state \cite{vp} 
\begin{eqnarray}
\sigma  &=&x\left| 00\right\rangle \left\langle 00\right| +(1-x)\left|
11\right\rangle \left\langle 11\right|   \nonumber \\
&&+\alpha \left| 00\right\rangle \left\langle 11\right| +\alpha ^{*}\left|
11\right\rangle \left\langle 00\right|   \label{q}
\end{eqnarray}
is given by 
\begin{eqnarray}
Er(\sigma ) &=&-x\ln x-(1-x)\ln (1-x)  \nonumber \\
&&+\lambda \ln \lambda +(1-\lambda )\ln (1-\lambda )  \label{r}
\end{eqnarray}
where $\lambda =\frac 12\{1+\sqrt{1-4[x(1-x)-|\alpha |^2]}\}$.

Other applications of this result can be found in ref. \cite{wz}.

We thank Prof. E. M. Rains for pointing out an important reference.

\widetext

\end{document}